\input{aipcheck}

\documentclass[final]{aipproc}

\layoutstyle{6x9}

\begin{document}

\title{The RHESSI Satellite and Classes of Gamma-ray Bursts}


\author{Jakub \v{R}\'{i}pa}{address={Astronomical Institute of the Charles University, V Hole\v{s}ovi\v{c}k\'{a}ch 2, Prague, Czech Republic}}

\author{Attila M\'{e}sz\'{a}ros}{address={Astronomical Institute of the Charles University, V Hole\v{s}ovi\v{c}k\'{a}ch 2, Prague, Czech Republic}}

\author{Ren\'{e} Hudec}{address={Astronomical Institute, Academy of Sciences of the Czech Republic, Ond\v{r}ejov, Czech Republic}}

\author{Claudia Wigger}{address={Paul Scherrer Institut, Villigen, Switzerland}}

\author{Wojtek Hajdas}{address={Paul Scherrer Institut, Villigen, Switzerland}}

\begin{abstract}
Some articles based on the BATSE gamma-ray burst (GRB) catalog claim the existence of a third population of GRBs, besides long and short. In this contribution we wanted to verify these claims with an independent data source, namely the RHESSI GRB catalog. Our verification is based on the statistical analysis of duration and hardness ratio of GRBs. The result is that there is no significant third group of GRBs in our RHESSI GRB data-set.
\end{abstract}

\maketitle

\section{Instrument}

 The Ramaty High Energy Solar Spectroscopic Imager (RHESSI) is a NASA Small Explorer satellite designed to study hard X-rays and gamma-rays from solar flares \cite{Lin2002}. It consists mainly of an imaging tube and a spectrometer. The spectrometer consists of nine germanium detectors (7.1 cm in diameter and a height of 8.5 cm) \cite{Smith2002}. They are only lightly shielded, thus making RHESSI also very useful to detect non solar photons from any direction. The energy range for GRB detection extends from about 50 keV up to 17 MeV depending on the direction. Energy and time resolutions are excellent for time resolved spectroscopy: $\Delta E$~=~3~keV at 1000 keV, $\Delta t$~=~1~$\mu$s. The effective area for near axis direction of incoming photons reaches up to 200 cm$^2$ at 200 keV. With a field of view of about half of the sky, RHESSI observes about one or two gamma-ray bursts per week. Data are stored event-by-event in onboard memory.

\section{Data Sample}

 The RHESSI GRB calalog (\url{http://grb.web.psi.ch}) contains 349 GRBs between the 14$^{th}$ February 2002 and the 8$^{th}$ May 2007. For deeper analysis we have chosen to use the subset of 332 GRBs with a signal-to-noise ratio better than 6. We have used the SolarSoftWare (\url{http://www.lmsal.com/solarsoft}) program under the Interactive Data Language (IDL) programming application as well as our own IDL routines to derive count light-curves and count fluences. We now take into account the fact that about 30 \% of the observed bursts were decimated. Decimation means that the count rate is sometimes automatically decreased for some detectors at some energies in order to save onboard memory. It biases the measured spectral hardnesses and light-curves. We have made software corrections to eliminate this effect and thus to obtain unbiased data.

\section{Duration Distribution}

Here we present the distribution of durations $T_{90}$. The $T_{90}$ is the time interval during which the cumulative counts increase from 5 \% to 95 \% above the background. Originally it was found (results from BATSE, Konus-Wind etc. instruments \cite{Kouveliotou1993,Aptekar1998}) that there exist two subclasses. The short one with $T_{90}$ < 2 s and the long one with $T_{90}$ > 2 s. However, some articles point to the existence of three subclasses of GRBs in the BATSE database with respect to their durations \cite{Horvath1998,Horvath2002,Horvath1999,Horvath2003}. Also some articles say that the third subclass (with intermediate duration) observed by BATSE is only a deviation caused by an instrumental effect \cite{Hakkila2000,Hakkila2004}. Therefore we decided to investigate this in the RHESSI database.

We have fitted one, two and three lognormal functions (Figure 1) and we have used the $\chi^2$ test to evaluate these fits. We have obtained a distribution with two maxima: one at approximately 0.32 s and one at approximately 17 s. From the statistical point of view, a single lognormal does not fit the observed distribution. For the best fitting with one lognormal function we obtained a confidence level lesser than 0.01 \%. Therefore the assumption that there is only one subclass can be rejected. For the best fitting with two lognormal functions we obtained a confidence level of 20.9 \%. For the best fitting with three lognormal functions we obtained a confidence level of 45.8 \%. A further result is that 18.3 \% of all RHESSI GRBs are short ones.
The question is whether the improvement of confidence level is statistically significant. To answer this question, we have used the F-test, as it is described in reference \cite{Band1997}. We obtained a critical value of $F_{0}$ = 2.42, which implies a probability of $P$($F$>$F_{0}$) = 12.1 \% that the improvement is just a statistical fluctuation. The value is too large to reject the hypothesis that the improvement in $\chi^2$ is only a statistical fluctuation.

\begin{figure}
  \includegraphics[width=1\textwidth]{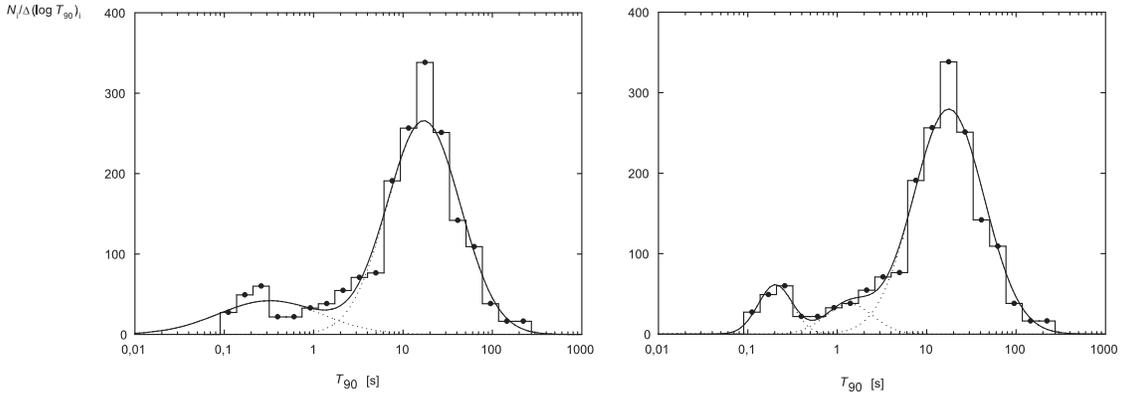}
  \caption{Distribution of the durations of bursts with two and three lognormal fits.}
\end{figure}

\section{Hardness Ratio vs. Duration}

Paper \cite{Mukherjee1998} has pointed to the existence of three GRB's subclasses in multiparameter space. Articles \cite{Horvath2004,Horvath2006} have claimed that in a 2D plane of hardness ratio vs. $T_{90}$ of the BATSE data-set, three subclasses of GRBs can be found. The hardness ratio is defined as a ratio of two fluences $F$ in two different energy bands integrated over the time interval $T_{90}$. These authors suggested that the third subclass has typical durations of about 5 s and the softest hardness ratio which is anti-correlated with the duration.
Here we present discussion of such possible distribution for the RHESSI data-set. Specifically we have used three energy bands: 25 - 120 keV, 120 - 400 keV and 400 - 1500 keV, and corresponding raw fluences therein: $F_1$, $F_2$ and $F_3$. In Figure 2, we show the hardness ratio $H_{231}$ = ($F_2$+$F_3$)/$F_1$ vs. $T_{90}$ with the best fit of two bivariate lognormal functions. For the estimation of the best fit, we have used the maximum likelihood method (see \cite{Horvath2002,Horvath2004,Horvath2006} and references therein). From our results it is seen that the observed distribution can be sufficiently described by only two groups. Points outside of the confidence level ellipses do not make any alone cluster.

\begin{figure}
  \includegraphics[height=.33\textheight]{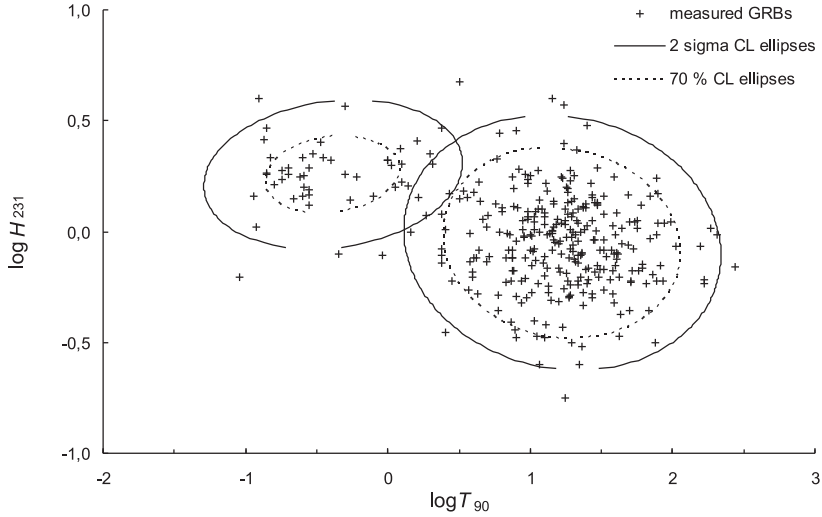}
  \caption{Hardness ratio vs. duration plot with the best fit of two bivariate lognormal functions.}
\end{figure}

\section{Conclusion}

The result of the lognormal fits to the duration distribution indicates that three lognormals fit are slightly better than two. However the improvement achieved by the addition of a third function is not statistically significant. There is no significant intermediate subgroup of GRBs in Figure 1. Also the two-dimensional hardness vs duration plot (Figure 2) does not demonstrate any remarkable third subgroup (with intermediate duration and soft hardness) in our RHESSI data sample.

\begin{theacknowledgments}

This research was partly supported by GAUK grant No. 46307. A. M. acknowledges the supports from the OTKA grant No. T48870 and from the Research Program MSM0021620860 of the Czech Ministry of Education.

\end{theacknowledgments}

\bibliography{mybib.bib}{}

\begin{thebibliography}{14}
\expandafter\ifx\csname natexlab\endcsname\relax\def\natexlab#1{#1}\fi
\providecommand{\enquote}[1]{``#1''}
\expandafter\ifx\csname url\endcsname\relax
  \def\url#1{\texttt{#1}}\fi
\expandafter\ifx\csname urlprefix\endcsname\relax\def\urlprefix{URL }\fi
\providecommand{\eprint}[2][]{\url{#2}}

\bibitem[{Lin} et~al.(2002)]{Lin2002}
R.~P. {Lin}, et~al., \emph{Solar Phys.} \textbf{210}, 3 (2002).

\bibitem[{Smith} et~al.(2002)]{Smith2002}
D.~M. {Smith}, et~al., \emph{Solar Phys.} \textbf{210}, 33 (2002).

\bibitem[{Kouveliotou} et~al.(1993)]{Kouveliotou1993}
C.~{Kouveliotou}, et~al., \emph{ApJ} \textbf{413}, 101 (1993).

\bibitem[{Aptekar} et~al.(1998)]{Aptekar1998}
R.~L. {Aptekar}, et~al., \emph{AIP Conf. Proc. of 4$^{th}$ Huntsville
  Symposium} p.~10 (1998).

\bibitem[Horv\'{a}th(1998)]{Horvath1998}
I.~Horv\'{a}th, \emph{ApJ} \textbf{508}, 757 (1998).

\bibitem[Horv\'{a}th(2002)]{Horvath2002}
I.~Horv\'{a}th, \emph{A\&A} \textbf{392}, 791 (2002).

\bibitem[{Horv\'{a}th} et~al.(1999)]{Horvath1999}
I.~{Horv\'{a}th}, et~al., \emph{AIP Conf. Proc. of 5$^{th}$ Huntsville
  Symposium} \textbf{526}, 200 (1999).

\bibitem[Horv\'{a}th(2003)]{Horvath2003}
I.~Horv\'{a}th, \emph{Statistical Challenges in Modern Astronomy III},
  Springer, 2003, p. 439.

\bibitem[{Hakkila} et~al.(2000)]{Hakkila2000}
J.~{Hakkila}, et~al., \emph{ApJ} \textbf{538}, 165 (2000).

\bibitem[{Hakkila} et~al.(2004)]{Hakkila2004}
J.~{Hakkila}, et~al., \emph{Baltic Astronomy} \textbf{13}, 211 (2004).

\bibitem[{Band} et~al.(1997)]{Band1997}
D.~L. {Band}, et~al., \emph{ApJ} \textbf{485}, 747 (1997), appendix A.

\bibitem[{Mukherjee} et~al.(1998)]{Mukherjee1998}
S.~{Mukherjee}, et~al., \emph{ApJ} \textbf{508}, 314 (1998).

\bibitem[{Horv\'{a}th} et~al.(2004)]{Horvath2004}
I.~{Horv\'{a}th}, et~al., \emph{Baltic Astronomy} \textbf{13}, 217 (2004).

\bibitem[{Horv\'{a}th} et~al.(2006)]{Horvath2006}
I.~{Horv\'{a}th}, et~al., \emph{A\&A} \textbf{447}, 23 (2006).

\end{thebibliography}
\bibliographystyle{aipproc}

\end{document}